\documentclass[twocolumn,aps,pra,floatfix,showpacs,preprintnumbers,groupedaddress]{revtex4}
\usepackage{amsmath}
\usepackage{graphicx}
\usepackage{amssymb}
\usepackage{bbm}
\usepackage{amsthm}
\usepackage{times}

\input{Qcircuit}

\newtheorem{theorem}{Theorem}
\newtheorem{definition}{Definition}
\newtheorem{lemma}{Lemma}

\newcommand{\one}{\ensuremath{\mathbbm{1}}}

\begin{document}

\title{Local Unitary Quantum Cellular Automata}

\author{Carlos A. P\'{e}rez-Delgado}

\affiliation{Institute for Quantum Computing, University of Waterloo, Waterloo,
ON N2L 3G1, Canada}

\author{Donny Cheung}

\affiliation{Institute for Quantum Computing, University of Waterloo, Waterloo,
ON N2L 3G1, Canada}

\begin{abstract}
In this paper we present a quantization of Cellular Automata. Our formalism is based on a lattice of \emph{qudits},
and an update rule consisting of local unitary operators that commute
with their own lattice translations. One purpose of this model is to act
as a theoretical model of quantum computation, similar to the quantum
circuit model. It is also shown to be an appropriate abstraction for
space-homogeneous quantum phenomena, such as quantum lattice gases,
spin chains and others. Some results that show the benefits of basing
the model on local unitary operators are shown: universality, strong
connections to the circuit model, simple implementation on quantum
hardware, and a wealth of applications.
\end{abstract}


\maketitle

\section{Introduction \label{sec:Introduction}}

The Cellular Automaton (CA) is a computational model that has been
studied for many decades \cite{neumann51,neumann66}. It is a
simple yet powerful model of computation that has been shown to
be Turing complete \cite{neumann66}. It is based on massive parallelism
and simple, locally constrained instructions, making it ideal for
various applications.  In particular, CA are very effective at simulating many classical physical systems,
including gas dispersion, fluids dynamics, ice formation, and even
biological colony growth \cite{cabook2}. 
Although usually simulated in software, CA hardware
implementations have also been developed. All of these characteristics
make CA a strong tool for moving from a physical system in nature,
to a mathematical model, to an implemented physical simulation.

More recently, the idea of \emph{Quantum} Cellular Automata (QCA)
has emerged. Several theoretical mathematical models have been proposed
\cite{grossing88,sw04,vandamthesis,watrous,cp05}. However, there is a lack
of applications developed within these models. On the other hand, \emph{ad hoc} models for specific applications
like quantum lattice gases \cite{meyer96,boghosian98}, among others
\cite{Forrester:2007gf}, have been developed. Several proposals
for scalable quantum computation (QC) have been developed that use
ideas and tools related to QCA \cite{Fitzsimons:2006ul,Vollbrecht:2006lq,benjamin1,Benjamin:2003oq,lloyd93}.  
Some of these  have been shown to be capable of universal
computation \cite{Raussendorf:2005cr,Shepherd:2006pd}. Other QCA tools have been used to solve, or propose solutions to,
particular problems in physics \cite{Imre:2006dq,Khatun:2005rr,cp06,Walus:2006qf,love-type2}.

However, there does not exist a comprehensive model of QCA that encompasses
these different views and techniques. Rather, each set of authors
defines QCA in their own particular fashion. In short, there is a
lack of a generally accepted QCA model that has all the attributes
of the CA model mentioned above: simple to describe; computationally
powerful and expressive; efficiently implemented in quantum software
\emph{and} hardware; and able to efficiently and effectively model
appropriate physical phenomena.

The purpose of this paper is to propose such a model. The model we
present here is based on intuitive and well-established ideas: qudits
as the basic building blocks (cells), and local unitary operators
as the basic evolution method  (local update rule).

The choice of local unitary operators as the basic evolution operator
ensures that the model is simple and easily explained to anyone familiar
with the field of quantum information. However, the choice is not
made merely for sake of simplicity: it provides us with an \emph{efficient
implementation} of QCA on quantum hardware, while still enjoying an
expressive richness strong enough to simulate any appropriate physical
system.

Formally, what we mean by efficient implementation is that there
exists a uniform family of quantum circuits that can each simulate
the evolution of a finite region of the QCA, for a specified number
of steps. Furthermore, we require that the \emph{depth} of each circuit
be strictly linear in the number of steps, and constant on the size
of the region being simulated. This last requirement is to ensure
that the QCA retains the quintessential quality of CA: \emph{massive
parallelism}.

We will refer to this formalization as the Local Unitary Quantum Cellular Automata (LUQCA) model, when we need to make the distinction  from other formal definitions of QCA.

In Section \ref{sec:Quantum-Circuits-and} we will see how \emph{any}
QCA properly defined in the model presented here can be efficiently implemented.
The fact that there is such a guarantee, without any further restraints,  
is one of the strongest features of the model herein presented. In
Section \ref{sec:Previous-QCA} we will see that in general, previous
models cannot make such a guarantee. We will also discuss what methods
 can be used to translate QCA in these models into the LUQCA model.

We will see in Sections \ref{sec:Quantum-Circuits-and} and \ref{sec:Modelling-Physical-Systems}
how insisting on efficient implementations does not at all limit
the expressive power of our QCA model. Section \ref{sec:Modelling-Physical-Systems}
will also show how most, if not all, physical systems of interest
with the proper characteristics---time and space homogeneity---can be
modeled using local unitary QCA. We will also prove computational
completeness in section \ref{sec:Quantum-Circuits-and}. Section \ref{sec:Previous-QCA}
discusses how valid QCA presented in other models can be rephrased in the
local unitary QCA scheme.

We begin in Section \ref{sec:Cellular-Automata} by briefly describing
classical CA in detail. Following that, we will endeavor to quantize
this model in the most natural way possible. The rest of the paper
presents results pertaining to the strengths of this model.

\section{Cellular Automata \label{sec:Cellular-Automata}}

In the classical model of cellular automata, we begin with a finite
set of states  $\Sigma$ and an infinite lattice of \emph{cells},
each of which is in one of the states in $\Sigma$. We have discrete
time steps, and at each time step $t$, the state of the lattice evolves
according to some rule. This rule gives the state of each cell at time
$t+1$ as a function of the states of the cells in its \emph{neighborhood},
which is simply a finite set of cells corresponding to a particular
cell. 

\begin{definition}[CA] A Cellular Automaton  is a 4-tuple $(L,\Sigma,\mathcal{N},f)$
consisting of a $d$-dimensional lattice of cells indexed by integers,
$L=\mathbb{Z}^{d}$, a finite set $\Sigma$ of cell states, a finite
neighborhood scheme $\mathcal{N}\subseteq\mathbb{Z}^{d}$, and a
local transition function $f:\Sigma^{\mathcal{N}}\rightarrow\Sigma$.
\end{definition}

The transition function $f$ simply takes, for each lattice cell position
$x\in L$, the states of the neighbors of $x$, which are the cells
indexed by the set $x+\mathcal{N}$ at the current time step $t\in\mathbb{Z}$
to determine the state of cell $x$ at time $t+1$. There are two
important properties of cellular automata that should be noted. First,
cellular automata are \emph{space-homogeneous}, in that the local
transition function performs the same function at each cell. Also,
cellular automata are \emph{time-homogeneous}, in that the local transition
function does not depend on the time step $t$.

We may also view the transition function as one which acts on the
entire lattice, rather than on individual cells. In this view, we
denote the state of the entire CA as a \emph{configuration} $C\in\Sigma^{L}$
which gives the state of each individual cell. This gives us a \emph{global}
transition function which is simply a function that maps $F:\Sigma^{L}\rightarrow\Sigma^{L}$.

\subsection{Reversible and Partitioned CA}

As a first step towards developing a theory of unitary CA we will revisit the theory of classical reversible automata.

A CA is reversible if for any configuration $C\in\Sigma^{L}$, and
time step $t\in\mathbb{Z}$ there exists a unique predecessor configuration
$C'$ such that $C=F(C',t)$. It is known that any Turing machine
can be simulated using a reversible CA \cite{toffoliphd}, so no
computational power is lost by this restriction.

One method that is used to construct reversible cellular automata
is \emph{partitioning}. In a partitioned CA, the transition
function is composed of local, reversible operations on individual
units of a partition of the lattice.

In order to formally define partitioned CA, we must expand the
definition of cellular automata, as partitioned CA are neither time-homogeneous
nor space-homogeneous in general. They are, however, periodic in both
space and time, and thus we set both a time period
 $T\geq1$ and a space period, given as a $d$-dimensional sublattice
$S$ of $L=\mathbb{Z}^{d}$. The sublattice $S$ can be defined using
a set $\{v_{k}:k=1,\ldots,d\}$ of $d$ linearly independent vectors
from $L=\mathbb{Z}^{d}$ as: \[
S=\left\{ \sum_{k=1}^{d}a_{k}v_{k}:a_{k}\in\mathbb{Z}\right\} .\]

\begin{definition} For a given fixed sublattice $S\subseteq\mathbb{Z}^{d}$,
we define a block $B\subseteq\mathbb{Z}^{d}$ as a finite subset
of $\mathbb{Z}^{d}$ such that $(B+s_{1})\cap(B+s_{2})=\emptyset$
for any $s_{1},s_{2}\in S$ with $s_{1}\neq s_{2}$, and such that
\[
\bigcup_{s\in S}(B+s)=\mathbb{Z}^{d}.\]
 \end{definition}

The main idea of the partitioned CA is that at different time steps,
we act on a different block partition of the lattice. We are now
ready to formally define the partitioned CA.

\begin{definition} A Partitioned CA is a 6-tuple $(L,S,T,\Sigma,\mathbf{B},\mathcal{F})$
consisting of 
\begin{enumerate}
\item a $d$-dimensional lattice of cells indexed by integers,
$L=\mathbb{Z}^{d}$;
\item a $d$-dimensional sublattice $S\subseteq L$;
\item a time period $T\geq1$;
\item a finite set $\Sigma$ of cell states;
\item a
block scheme $\mathbf{B}$, which is a sequence $\{B_{0},B_{1},\ldots,B_{T-1}\}$
consisting of $T$ blocks relative to the sublattice $S$; and
\item a local
transition function scheme $\mathcal{F}$, which is a set $\{f_{0},f_{1},\ldots,f_{T-1}\}$
of reversible local transition functions which map $f_{t}:\Sigma^{B_{t}}\rightarrow\Sigma^{B_{t}}$.
\end{enumerate}
\end{definition}

At time step $t+kT$ for $0\leq t<T$ and $k\in\mathbb{Z}$, we perform
$f_{t}$ on every block $B_{t}+s$, where $s\in S$. In order to find
the reverse of a partitioned CA, we simply give the reverse block
scheme, $\mathbf{B}=\{B_{T-1},\ldots,B_{1},B_{0}\}$, and the reverse
function scheme, $\mathcal{F}=\{f_{T-1}^{-1},\ldots,f_{1}^{-1},f_{0}^{-1}\}$.

\begin{figure}
\begin{center}
\includegraphics{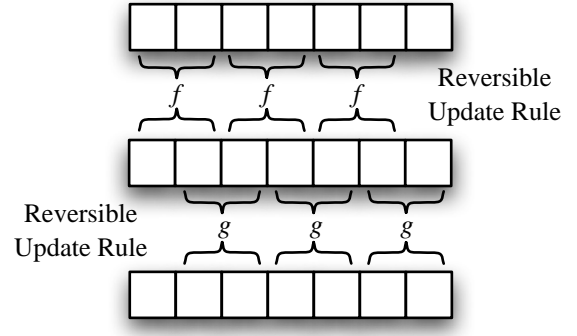}
\end{center}
\caption{\emph{Partitioned Cellular Automaton}}
\end{figure}

Although the partitioned CA is not time- or space-homogeneous, it
can be converted into a regular CA, on the lattice $S$ (which is
isomorphic to $\mathbb{Z}^{d}$), with cell states $\Sigma^{B}$,
where the new local transition function simulates $T$ time steps
of the partitioned CA in one time step.

In the original partitioned CA scheme as described by Margolus \cite{cabook},
the sublattice was fixed as $S=2\mathbb{Z}^{d}$, and the block scheme
was fixed with two partitions: $B_{0}=\{(x_{1},x_{2}\ldots,x_{d}):0\leq x_{j}\leq1\}$
and $B_{1}=\{(x_{1},x_{2}\ldots,x_{d}):1\leq x_{j}\leq2\}$.

\section{Local Unitary QCA \label{sec:Local-Unitary-QCA}}

Now, with a formal notion of CA, we can proceed to give a \emph{quantization}.
As mentioned earlier, we will have very specific goals in mind.

\subsection{Model Requirements}

First, we want to develop an intuitive model that is both simple to
work with, and to develop algorithms for. At the same time, we want
this model to be an obvious \emph{extension} of classical CA and
to reduce to classical CA behavior under reasonable limits.

Second, we want to keep our model grounded in physical realities. This
has a couple of strong consequences. Namely, we approach CA, even
classical CA, not as abstract mathematical structures, but as models
representing real physical systems. As a consequence, we expect our model to reliably model quantum
systems with appropriate behavior, \emph{e.g.}, spin chains.
 Also, an algorithm described
in our model should be easy to  translate to an actual physical implementation
on such quantum systems. We show in Section \ref{sec:Quantum-Computation}
that this is so.

\subsection{A First Approach \label{sub:Shift-Right}}

The first step in our quantization of CA is to change the state space
of a cell to reflect a quantum system. There are several methods for
doing so, however we believe that the most natural way to approach
this is to convert the alphabet of the cellular automaton, $\Sigma$,
into orthogonal basis states of a Hilbert space, $\mathcal{H}_{\Sigma}$.
Formally, every cell $x\in L$ is assigned a \emph{qudit}, $\ket{x}\in\mathcal{H}_{\Sigma}$.
This gives us a strong intuitive tool, as the notion of a lattice
of qudits should be familiar to anyone working in quantum information
theory.

As we shall see, our approach is also physically grounded, in that it is possible
to describe this model in terms of a quantum system evolving according
to a Hamiltonian. As an example, spin chains can be directly described
by such mathematical constructions. Lattice gases, though not originally
modeled in this way, can also be easily described by such mathematical
constructs. Perhaps the most obvious physical example is the pulse-driven
quantum computer.

We also wish to quantize the standard classical CA update rule. However,
this process cannot necessarily proceed in the most obvious manner.
In a classical CA, every cell is instantaneously updated in parallel.
We wish to replace this classical cell update rule with a quantum
analogue that acts appropriately on the qudit lattice described above.
For a quantum unitary operation to act as a quantum cell update rule,
this operator needs to fulfill the following two restrictions:

\begin{enumerate}
\item The operator must act on a finite subset of the lattice. Precisely
$U_{x}:\mathcal{H}(\mathcal{N}_{x})\rightarrow\mathcal{H}(\mathcal{N}_{x})$
where $\mathcal{N}_{x}=\mathcal{N}+x\subseteq L$ is the finite neighborhood
about the cell $x$. 
\item The operator must commute with lattice translations of itself. Precisely,
we require that $[U_{x},U_{y}]=0$ for all $x,y\in\mathbb{Z}^{n}$.
\end{enumerate}
The first condition is an immediate condition for any rule, quantum
or otherwise, to qualify as a CA update rule. The second condition
allows the operators $U_{x}$ for $x\in\mathbb{Z}^{n}$ to be applied
in parallel without the need to consider any special or particular 
ordering of the operators.

It should be clear that any evolution defined in such manner represents
a valid quantum evolution which can be ascribed to some physical system.
The global evolution of the lattice can be described as \[
U=\prod_{x}U_{x},\]
 which is well-defined, due to the two conditions given above.

The question that remains is whether this model properly describes
what we intuitively would regard as QCA. Properly, there are two questions:

\begin{enumerate}
\item Can all entities described by the model above be properly classified
as QCA? 
\item Can all systems that are identified as QCA be properly described in
the model above? 
\end{enumerate}
The answer to the first question is \emph{yes}, since the update rules
are local and can be applied in unison throughout the lattice. Also,
the global unitary operator for the evolution of the lattice is properly
defined and space-homogeneous, as desired.

The answer to the second question is, unfortunately, \emph{no}. We
now present a simple system that one might consider to be a valid QCA,
but cannot be described in the above model.

The counterexample is as follows. We start with a one-dimensional lattice
of qudits. For each lattice cell $x\in L$, we associate with it a
quantum state $\ket{\psi_{x}} \in\mathcal{H}_{\Sigma}$. Although
in general, the configuration of a QCA may not be separable with respect
to each cell, the configuration can still be described in terms of
a linear superposition of these separable configurations. Thus, it
suffices to consider such configurations.

At each time step we wish to have every value shifted one cell to
the right. In other words, after the first update each cell $x$ should
now store the state $|\psi_{x-1}\rangle$. After $k$ steps each cell
$x$ should contain the state $|\psi_{x-k}\rangle$. In fact, such
a transition function cannot be implemented by \emph{any} local unitary
process.

To see why this is so, suppose that we had a transition function $f$,
which is the product of a finite number of operations, $f=f_{n}f_{n-1}\ldots f_{1}$,
such that each operator $f_{j}$ is the (potentially infinite) product
of local unitary operators over disjoint neighborhoods. Note that
this gives us the most general description possible of a depth-$n$
quantum circuit implementation of this linear QCA using only local
unitary operators. Now, consider an individual cell, $x_{0}$. By
analyzing the dependencies of the individual local unitary operators
which make up the transition function $f$, it is possible to find
a range of cells, $P=\{x:a\leq x\leq b\}$ for some $a,b\in\mathbb{Z}$
such that $x_{0}\in P$, and the value of the quantum state at cell
$x_{0}$ after the application of the transition function depends
only on the cells of $P$.

We now divide the transition function $f$ into two functions, $f=hg$,
where $g$ applies sufficiently many local unitary operators from
$f$ over the cells of $P$ so that the new value of the quantum state
at cell $x_{0}$ is computed, without violating any of the dependency
relations from $f$. Then, $h$ simply applies the remainder of the
local unitary operators, as appropriate. Note that since $g$ necessarily
contains any local unitary operators from $f$ which operate on the
cell $x_{0}$, the operation $h$ does not. Since $h$ does not perform
any operation between cells $x<x_{0}$ and cells $x>x_{0}$, in order
to implement the shift-right transition function, the cells $\{x:x_{0}<x\leq b\}$
must contain enough quantum information after $g$ has been applied
to reconstruct the information in the cells $\{x:x_{0}\leq x\leq b\}$.
This is clearly impossible.

In order to resolve this issue, we need to analyze the classical CA
parallel update rules more closely.
In the classical CA, the local update rule for a given cell reads
the value of the cell, and the values of its neighboring cells. It
performs a computation based on these values, and then updates the
cell's value accordingly. Herein lies the problem: \emph{read} and
\emph{update} are modeled in a classical CA as a single atomic action
that can be applied throughout the lattice in parallel simultaneously.
However, in a physical setting, these two operations cannot be implemented
in this manner. When simulating CA in classical computer architectures,
the canonical solution is to use two lattices in memory: one to store
the current value, and one to store the computed updated value. Even
if we consider hardware implementations of CA, these need to keep
the values of the inputs to the transition function while this function
is being calculated.

The formal CA model does not need to consider this implementation
detail, as it is a mathematical construction and has no claims to
directly model a physical system implementing a CA. When developing
a QCA model, one cannot take the same liberty. The name itself, QCA,
includes reference to an underlying \emph{quantum} physical reality.
It is our intention that this model faithfully, if abstractly, represents
real physical systems. Although there is some value in mathematical
constructions which do not correspond directly to any physical systems,
this is not the goal of the constructions presented in this paper.

\subsection{A New Approach}

We now make an adjustment to our QCA model, given the importance
of maintaining independent \emph{read} and \emph{update} operations.
Instead of having one unitary operator replacing the single atomic
operation in the CA model, we define our QCA update rule as consisting
of two unitary operators. The first operator, corresponding to the
\emph{read} operation, will be as defined above: a unitary operator
$U_{x}$, $x\in L$ acting on the neighborhood $\mathcal{N}_{x}$,
which commutes with all lattice translations of itself, $U_{y}$,
$y\in L$. The second operator, $V_{x}$, $x\in L$, corresponds to
the \emph{update} operation, and will only act on the single cell
$x$ itself.

The intuition is as follows: in our physical model, instead of having
separate lattices for the \emph{read} and \emph{update} functions,
we expand each lattice cell to also contain any space resources necessary
for computing the updated value of the cell. The operator $U_{x}$
reads the values of the neighborhood $\mathcal{N}_{x}$, performs
a computation, and stores the new value in such a way that does not
prevent neighboring operators $U_{y}$ from correctly reading its
own input values. This allows each cell to be operated upon independently,
in parallel, without any underlying assumptions of synchronization.
After all the operations $U_{x}$ have been performed, the second
unitary $V_{x}$ performs the actual update of the lattice cell.

With this new model for the update operation, we can again approach
the two questions given above as to whether this model adequately
describes what we might intuitively regard as QCA.

First, it is clear that all entities described by this updated model
can still be properly classified as QCA. The local update rule $R_{x}=V_{x}U_{x}$
is still a valid quantum unitary operation, and the global update
rule \[
R=VU=\left(\bigotimes_{x}V_{x}\right)\left(\prod_{x}U_{x}\right)\]
 is both well-defined and space-homogeneous.

Now, in order to properly investigate whether all physical systems
which can be described as QCA can be described within this new model,
it is necessary to verify the following:

We must first compare our model to existing CA models, both classical
and quantum, in order to ensure that our model subsumes all proper
CA described in these models. Secondly, we must also show that any
known physical system which behaves according to quantum mechanics
and satisfies the CA preconditions of being driven by a local, space-homogeneous
interaction can be described by our model.

As an example, the qubit shift-right QCA mentioned above can now be
described in this model, by including ancillary computation space
with each lattice cell.

We will tackle this question in more depth in the upcoming sections.
First, we present a formal definition of the QCA model which we will
adopt, as described in this section.

\begin{definition}[QCA] A Quantum Cellular Automaton  is a 5-tuple
$(L,\Sigma,\mathcal{N},U_{0},V_{0})$ consisting of 
\begin{enumerate}
\item a $d$-dimensional
lattice of cells indexed by integers, $L=\mathbb{Z}^{d}$,
\item a finite
set $\Sigma$ of orthogonal basis states,
\item   a finite neighborhood scheme
$\mathcal{N}\subseteq\mathbb{Z}^{d}$,
\item a local read function $U_{0}:\left(\mathcal{H}_{\Sigma}\right)^{\otimes\mathcal{N}}\rightarrow\left(\mathcal{H}_{\Sigma}\right)^{\otimes\mathcal{N}}$,
and 
\item a local update function $V_{0}:\mathcal{H}_{\Sigma}\rightarrow\mathcal{H}_{\Sigma}$.
\end{enumerate}
The \emph{read} operation carries the further restriction that any
two lattice translations $U_{x}$ and $U_{y}$ must commute for all
$x,y\in L$. \end{definition}

Each cell has a finite Hilbert space associated with it $\mathcal{H}_{\Sigma}=\textrm{span}\left(\{|x\rangle\}_{x\in\Sigma}\right)$.
The reduced state of each cell $\rho_{x}$ is a density operator over
this Hilbert space.

The initial state of the QCA is defined in the following way. Let
$f$ be any computable function that maps lattice vectors to pure
quantum states in $\left(\mathcal{H}_{\Sigma}\right)^{\otimes k^{d}}$,
where $d$ is the dimension of the QCA lattice, and $k$ is the length of a side of a $d$-dimensional hypercube, which we use to define blocks that are initialized to particular states.
 Then for any lattice vector $\mathbf{z}=(z_1k,z_2k,\ldots,z_dk) \in \mathbb{Z}^d$ the initial state of the lattice hypercube delimited
by $\left(z_1k,z_2k,\ldots,z_dk\right)$ and $\left((z_1+1)k-1,\right.$ $\left. (z_2+1)k-1,\ldots,(z_d+1)k-1\right)$
is set to $f(\mathbf{z})$.

Intuitively, each block represents a volume of the QCA that is initialized to a particular pure state. Each block is initialized independently.
In particular, $f$ can have a block size of one cell, initializing
every cell to the same state in $\Sigma$. It can also have more complicated
forms such as having every pair of cells in a one dimensional QCA initialized
to some maximally entangled state. Particularly useful are functions $f$ that initialize a finite region about the origin to some interesting state---the input of the QCA---and the rest of the lattice to some quiescent state (see below).

The local update rule acting on a cell $x$ consists of the operation
$U_{x}$ followed by the single-cell operation $V_{x}$. Both $U_{x}$
and $V_{x}$ are restricted to being computable unitary operators.
The global evolution operator $R$ is as previously defined.

\subsection{Quiescent States}

Our QCA definition follows the classical CA convention in defining
the model over an \emph{infinite} lattice. However, we will often
be concerned only with finite regions of the QCA. One reason, for example, is that any
physical implementation of a QCA using quantum hardware will, by necessity,
simulate only a finite region of the QCA. Another reason is for simulating
physical phenomena. For instance, in Section \ref{sec:Modelling-Physical-Systems},
we will be interested in simulating \emph{finite size} chains of spin-$\frac{1}{2}$
particles.

Sometimes, it can be appropriate to simply use finite QCA with cyclic
boundary conditions. In this case, we envision the lattice as a closed
torus. This is a standard and well-known practice with CA. For example,
we can use this technique if the spin chain we wish to simulate is
\emph{closed}, that is, it itself wraps around. For other applications,
this will not be appropriate, for example, when trying to simulate
an \emph{open} spin chain. This is a chain which does \emph{not} wrap
around, but rather has two distinct end points. Another example will
be the spin-signal amplification algorithm in Section \ref{sec:Algorithms},
which uses a finite size cube ancilla system.

In such cases, the most appropriate way to proceed is to make use
of a \emph{quiescent} state, which is a cell state that is guaranteed
to remain invariant under the update rule, regardless of the states
of its neighbors.
For instance, in the case of the finite spin-$\frac{1}{2}$ chains,
we can use three state cells. We use the state labels $\ket{{+}1}$
and $\ket{{-}1}$ to refer to the presence of a spin-$\frac{1}{2}$
particle in a given cell position in the states $\frac{1}{2}(\one+\sigma_{z})$
and $\frac{1}{2}(\one-\sigma_{z})$ respectively. A third state, labeled
$\ket{0}$ denotes the absence of any particle in that cell location.
One  then need only ensure that the update rule correctly acts on
states $\ket{{+}1}$ and $\ket{{-}1}$, while leaving state $\ket{0}$
unaffected.

\begin{figure}
\begin{center}
\includegraphics{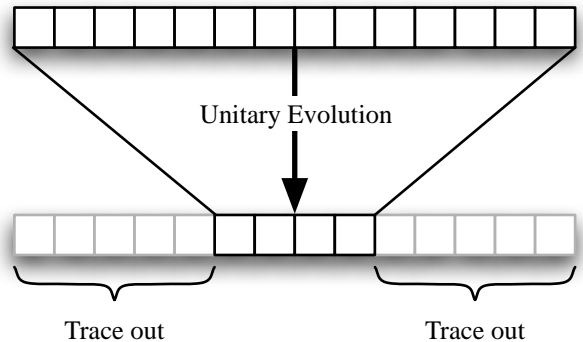}
\end{center}
\caption{\emph{Past lightcone of a region} $S$: This represents a one-dimensional
local unitary QCA. In order to obtain the state of the region of interest,
the dark region at the bottom, one must consider not just the region
itself, but anything that might affect the state of the region with
the course of the simulation: its past lightcone. One may then trace
out the unneeded regions. \label{fig:lightcone}}
\end{figure}

Quiescent states are also very useful for the purposes of \emph{simulation},
and physical implementation. Normally, if one is interested in the
state of a region $S$ of the lattice after $k$ steps of the QCA
update rule, one would need to look at the \emph{past lightcone} of
$S$. If the local update rule has a neighborhood of radius $r$,
then one needs to include $kr$ additional cells in each direction
beyond the border of $S$. This is because any information in the
past lightcone of $S$ has the ability to affect cells within $S$,
as shown in Figure \ref{fig:lightcone}. Note that since the size
of the region needed by the simulation is determined by the number
of time steps of the QCA we wish to simulate, one needs to fix the
number of steps in the simulation beforehand. However, if a given
QCA has a quiescent state, and all cells outside the finite region
under consideration are initialized to this quiescent state, then the
simulation of this QCA need only include this region for any number
of simulated time steps.

\section{Quantum Circuits and Universality \label{sec:Quantum-Circuits-and}}

In this section we explore two important aspects of the QCA model
we introduced in Section \ref{sec:Local-Unitary-QCA}. These aspects
relate to QCA as a model of computation. First, it is important to
show that QCA are capable of universal quantum computation. We demonstrate
this using a simulation of an arbitrary quantum circuit using a two-dimensional
QCA.

We also show that any QCA can be simulated using families of quantum
circuits. A quantum circuit is defined as a finite set of gates acting
on a finite input. One can then define a \emph{uniform} family of
quantum circuits, with parameters $S$ and $t$, such that each circuit
simulates the finite region $S$ of the QCA for $t$ update steps.
By uniformity we mean that that there exists an effective procedure,
such as a Turing machine, that on input $(S,t)$ outputs the correct
circuit.

We will show that our simulation is \emph{efficient}, as defined in
Section \ref{sec:Introduction}. Specifically, in order to simulate
a QCA on a given region, for a fixed number of time steps, we give
a quantum circuit simulation with a depth which is linear with respect
to the number of time steps, and constant with respect to the size
of the simulated region.

\subsection{Simulation of QCA by Quantum Circuits}

We begin by showing the latter of the two results described above.
We proceed incrementally, showing first how to produce a quantum circuit
that can simulate a single update step of a simple QCA.

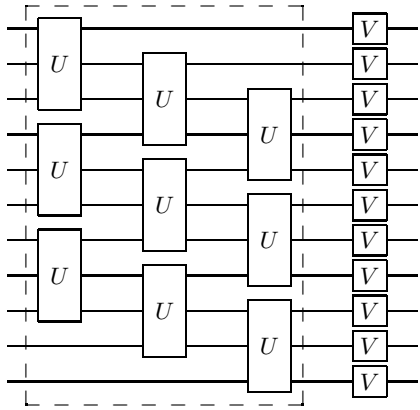
\begin{figure}
\begin{align*}
\Qcircuit @C=1.3em @R=0.2em {
& \multigate{2}{U} & \qw & \qw & \qw & \qw & \qw & \gate{V} & \qw \\
& \ghost{U} & \qw & \multigate{2}{U} & \qw & \qw & \qw & \gate{V} & \qw \\
& \ghost{U} & \qw & \ghost{U} & \qw & \multigate{2}{U} & \qw  &\gate{V} & \qw  \\
& \multigate{2}{U} & \qw & \ghost{U} & \qw  & \ghost{U} & \qw & \gate{V} & \qw \\
& \ghost{U} & \qw & \multigate{2}{U} & \qw  & \ghost{U} & \qw & \gate{V} & \qw \\
& \ghost{U} & \qw & \ghost{U} & \qw & \multigate{2}{U} & \qw & \gate{V} & \qw \\
& \multigate{2}{U} & \qw & \ghost{U} & \qw  & \ghost{U} & \qw & \gate{V} & \qw \\
& \ghost{U} & \qw & \multigate{2}{U} & \qw  & \ghost{U} & \qw & \gate{V} & \qw \\
& \ghost{U} & \qw & \ghost{U} & \qw & \multigate{2}{U} & \qw & \gate{V} & \qw \\
& \qw & \qw &  \ghost{U} & \qw & \ghost{U} & \qw & \gate{V} & \qw \\
& \qw & \qw &  \qw & \qw & \ghost{U} & \qw & \gate{V} & \qw  \qw \gategroup{1}{2}{11}{6}{1em}{--}}
\end{align*}
\caption{\emph{Quantum circuit simulation of a QCA update step:} The dotted
area represents the read phase. A read operator $U$ must be applied
to each qubit, and its two neighbors. Since $U$ commutes with its
translations, we are at liberty to apply the $U$ operators in any order.
The update phase consists of the operator $V$ being applied to every
qubit. \label{fig:soundness}}
\end{figure}

\begin{lemma} Any finite region of a one-dimensional QCA with a symmetric
neighborhood of radius one, where cells are individual qubits, can
be simulated by a quantum circuit.

\begin{proof} The simulation of an individual update step of this
QCA is simple. Recall that the operators $U_{x}$, each acting on
3 qubits, all commute with each other. Therefore, the $U_{x}$ operators
may be applied in an arbitrary order. The operators $V_{x}$ can all
be applied to their respective qubits once all $U_{x}$ operators
have been applied. Figure \ref{fig:soundness} gives a visual representation
of this construction. In order to simulate an arbitrary number of
steps, we simply need to repeatedly apply the above construction.
 Finally,
although we represented the operators $U$ in our diagram as single,
three-qubit operators, to complete the simulation we decompose $U$
into an appropriate series of one and two qubit gates from a universal
gate set. \end{proof} 

\end{lemma}

In the case of one-dimensional QCA with a nearest neighbor scheme,
and cells consisting of one qubit, the operator $U$ is simply acting
on three qubits. Still this operator $U$ needs to be decomposed into
a series of one and two qubit gates $U=U_{n}U_{n-1}\ldots U_{2}U_{1}$
 taken from a set of universal gates.

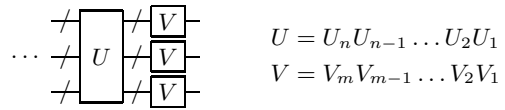
\begin{figure}
\begin{minipage}{0.4\columnwidth}%
\begin{align*}
\Qcircuit @C=0.6em @R=0.2em {
& {/} \qw & \multigate{2}{U} &  {/} \qw &  \gate{V} & \qw  \\
\lstick{\ldots} & {/}  \qw & \ghost{U} & {/}  \qw  &  \gate{V} &\qw \\
& {/}  \qw & \ghost{U} &  {/} \qw  &  \gate{V} & \qw  \\
}
\end{align*}%
\end{minipage}%
\begin{minipage}{0.4\columnwidth}%
\begin{align*}
U & = U_n U_{n-1} \ldots U_2 U_1 \\
V & = V_m V_{m-1} \ldots V_2 V_1
\end{align*}%
\end{minipage}%
\caption{\emph{Decomposition of general qudit gates}. \label{fig:quditwire}}
\end{figure}

In order to extend the construction to allow cells with qudits of
arbitrary dimension $d$, we first replace the single qubit wires
in Figure \ref{fig:soundness} with qudit wires as in Figure \ref{fig:quditwire}.
Then each gate $U_{x}$ and $V_{x}$ are decomposed into one and two
qubit gates as in the aforementioned figure. The same construction
technique works in order to deal with arbitrary dimensions, and arbitrary
cell neighborhood sizes.

Note that $m$ and $n$ are constants, determined by the structure
of the QCA. For very complicated QCA these constants can
be potentially large. However, once the QCA has been defined
these parameters are set, and hence do not asymptotically affect the
complexity of simulating a region of the QCA for a particular length
of time.

As our simulation above does not set a region size to be simulated,
any region size can be simulated with an appropriate construction.
An arbitrary number of time steps can be simulated by simply iterating
the above construction. With this in mind, as well as the previous
lemma, we can now state the following:

\begin{theorem} For every QCA $Q$ there exists a family of quantum
circuits, parameterized by $(S,t)$, each acting on $O(m\log|\Sigma|)$
inputs, and with circuit depth $O(t)$ which simulates a finite region
of $S$ of $Q$ consisting of $m = |S|$ cells, for $t$ time steps\end{theorem}

This is a very important result, as it demonstrates that the local
unitary QCA model does not admit automata which are somehow {}``not
physical''. More precisely, any behavior that can be described by
a QCA can be described by the more traditional quantum circuit model.
Furthermore, such descriptions retain the high parallelism inherent
to QCA.

\subsection{Simulation of Quantum Circuits by QCA}

Next, we show the converse result from the one above, thus showing
that local unitary QCA are capable of efficient universal quantum
computation.

\begin{theorem} There exists a universal QCA $Q_{u}$ that can simulate
any quantum circuit with at most a linear slowdown, 
by using an appropriately encoded initial state.
\begin{proof}
We proceed by constructing the QCA $Q_{u}$ over a two-dimensional
lattice. We will basically \emph{`draw'} the circuit onto the lattice.
The qubits will be arranged top to bottom, and the wires will be visualized
as going from left to right.

Each cell will consist of a number of fields, or registers. The cell
itself can be thought of as the tensor product of quantum systems corresponding
to these registers. 

The first register, the state register, consists
of a single qubit which corresponds directly to the value on one of
the wires of the quantum circuit at a particular point in the computation.
This value will be shifted towards the right as time moves forward.

Next is the gate register. This register will be initialized to a
value corresponding to a gate that is to be applied to the state register,
at the appropriate time.  

There is also  a clock register, which will keep
track of  the current time step of the simulation. There are two phases
to the simulation, an \emph{`operate'} and a \emph{`carry'} step. 

There
is finally a single qubit active register, that keeps a record of
which cells are currently actively involved in the computation. This
register is either set to \emph{true} or \emph{false}.

The local read operator $U_{x}$ proceeds as follows. The neighborhood scheme is the von Neumann neighborhood of radius one, \emph{i.e.} the cells directly above, below and to either side of the cell. The read operator acts non-trivially only on the one cell directly above, and the one directly to the
left. However, the bigger neighborhood is needed to ensure unitary evolution, and translation invariance.

If the clock register is set to \emph{operate}, then a quantum gate
is applied to the state register of the current cell (and possibly
the state register of the upwards neighbor). For this, we fix a finite
set of universal gates consisting of the controlled phase gate and
some set of single-qubit operators. The choice of the controlled phase gate, as opposed to say controlled not, is to ensure
that $U_{x}$ commutes with translations of itself. Any one-qubit
unitary gates that form a universal set will work.

If the clock register is set to \emph{carry}, then the state register
will be swapped with the state register of the left neighbor if and
only if the following conditions occur: the active register is set
to \emph{true} on the left neighbor, and set to false on the current
cell, and the clock register is set to carry on all the neighbors (above, below, and to either side).
These extra checks are required to ensure the operator $U_{x}$ commutes
with translations of itself.

\begin{figure}
\begin{center}
\includegraphics{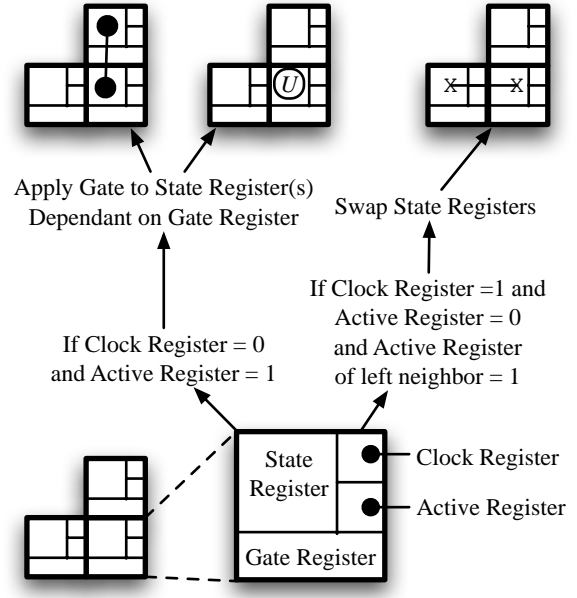} 
\end{center}
\caption{\emph{Universal QCA update rule.} \label{fig:universalQCA}}
\end{figure}

Figure \ref{fig:universalQCA} gives a visual representation of the
update rule operator $U_{x}$. Operator $V_{x}$ simply updates the
clock register, applying a $NOT$ gate at each time step.

Finally, the initial state is set as follows. There is one horizontal
row for each wire in the quantum circuit. Every column represents
a time step in the quantum circuit. The cells are initialized to have
their gate registers set to the appropriate gate, if there is a gate,
in the wire corresponding to its row, and in the time step corresponding
to its column. The clock register is set to operate, and the state register is set $\ket{0}$ initially on all cells. The first column of the quantum circuit is set to active, all other cells are set to inactive.

This construction  can only natively simulate circuits with
nearest-neighbor gates. In order to encode arbitrary circuits, it
is necessary to translate the circuit into one using only nearest-neighbor
gates by adding swap gates where needed. This is the cause of the worst-case linear slowdown, mentioned in the statement of this theorem.
\end{proof}
\end{theorem}

The previous result is important in that it proves that the QCA model
is computationally complete. It also gives a recipe for implementing
quantum circuit algorithms on two-dimensional QCA. It is important to mention
that it is also possible to show that one-dimensional LUQCA are universal
for quantum computing. For a complete proof see \cite{cpthesis}.

In the following sections, by showing how physical systems can implement
QCA, we complete a formula for implementing quantum algorithms on
physical systems using QCA methods. We will see, however, that the
strongest virtue of this QCA model lies not in its ability to simulate
quantum circuits. Rather, it lies in the algorithms that take natural
advantage of the QCA structure.

\section{Modelling Physical Systems \label{sec:Modelling-Physical-Systems}}

We stated before that one of our goals in developing a QCA formalism
is to create a useful modeling tool for quantum systems. Classical
CA are used for simulating various phenomena based on classical information,
such as sea ice formation, fluid dynamics, or voter systems \cite{cabook,cabook2}.
Similarly, we expect QCA to be able to model different types of physical
systems based on quantum information, with dynamics which are based
on time and space homogeneous local interactions.

Physical systems that fall in this category include Ising and Heisenberg
interaction spin chains, solid state NMR, and quantum lattice gases.
We will be looking at some of these systems in this section.

\subsection{Spin Chains}

Spin chains are perhaps the most obvious candidate for physical systems
being modeled with QCA. Indeed, Ising interaction spin chains, and
in general, any spin chain with a coupling Hamiltonian which commutes
with its own lattice translations can be implemented easily.

Suppose we have a linear spin chain of length $N$, indexed by $n\in\mathbb{Z}$.
Each spin $n$ is coupled to its nearest neighbor $n+1$, with a coupling
Hamiltonian $J\sigma_{z}^{(n)}\sigma_{z}^{(n+1)}$, where $J$ is
the coupling strength constant. Note that the coupling Hamiltonian
does commute with its lattice translations. The Hamiltonian for the
entire spin chain is: \[
H_{\textrm{I}}=\sum_{n=1}^{N-1}J\sigma_{z}^{(n)}\sigma_{z}^{(n+1)}.\]

It is a simple matter to give a discrete time approximation to such
a spin chain. First, we fix a time step interval $\Delta t$. Our
QCA model will allow for simulation of the spin chain for time steps
in multiples of $\Delta t$. Hence, while the choice of $\Delta t$
is arbitrary, it is important in determining the resolution of the
simulation.

For a simulation of the Ising spin chain, the QCA lattice consists
of a one-dimensional array, where each cell is a single qubit. The neighborhood
of each cell $n$ simply consists of the cell and its right neighbor
$n+1$. The local rule operator $U_{n}$ is given as: \[
U_{n}=e^{-iJ\sigma_{z}^{(n)}\sigma_{z}^{(n+1)}\Delta t}.\]

The operator $V_{n}$ is simply the identity operator. Note that the
operator $U_{n}$ commutes with its translations, that is, $[U_{n},U_{m}]=0$,
for all $n,m\in\mathbb{Z}$. Furthermore, the global operator \[
U=\prod_{n=1}^{N-1}U_{n}\]
 satisfies \[
U=e^{-iH_{\textrm{I}}\Delta t}.\]
Hence, the QCA construction faithfully simulates the Ising spin chain
for  times that are integer multiples of $\Delta t$, as desired.

A more complicated endeavor is to construct a QCA simulation of a
spin chain whose coupling Hamiltonians do not commute with each other.
In particular we examine the Heisenberg spin chain as an example.
Let the coupling Hamiltonian between spins $n$ and $n+1$ be \begin{eqnarray*}
H_{\textrm{H}}^{(n,n+1)} & = & J(\sigma_{x}^{(n)}\sigma_{x}^{(n+1)}+\sigma_{y}^{(n)}\sigma_{y}^{(n+1)}\\
 &  & \quad+\sigma_{z}^{(n)}\sigma_{z}^{(n+1)}-\one\otimes\one)\end{eqnarray*}

Here, note that $H_{\textrm{H}}^{(n,n+1)}$ does not commute with
its translations $H_{\textrm{H}}^{(m,m+1)}$. The Hamiltonian of the
total system is \[
H_{\textrm{H}}=\sum_{n=1}^{N-1}H_{\textrm{H}}^{(n,n+1)}\]

A QCA simulation of the Heisenberg spin chain presented above is still
possible, however, with the help of two powerful tools: \emph{Trotterization},
and \emph{cell coloring}. The first technique is well known in physics,
the second is a tool developed for QCA. Together, they allow for simulation
of complicated and almost arbitrary Hamiltonians by QCA.

Trotterization is a technique by which a Hamiltonian is approximated
using a combination of non-commuting Hamiltonians whose sum adds up
to the original Hamiltonian. In other words, it is possible to approximate
with bounded error the evolution due to the Hamiltonian $H=H_{a}+H_{b}$
by alternately evolving the system under the Hamiltonians $H_{a}$
and $H_{b}$ even if these two do not commute. Precisely, we can give
a first-order approximation \[
e^{-i(H_{a}+H_{b})\Delta t}=\left(e^{-iH_{a}\Delta t/k}e^{-iH_{b}\Delta t/k}\right)^{k}+\delta.\]

In the case that $\left\Vert [H_{a},H_{b}]\right\Vert \Delta t^{2}\ll1$,
the error $\delta$ is $O(\Delta t^{2}/k)$. Higher order techniques
can achieve error rates of $O(\Delta t^{m}+1/k^{m})$ at the cost
of using $O(2^{m})$ gates. Though the number of gates increases exponentially,
the time required for each gate \emph{decreases} exponentially as
well.

In the case of our QCA simulation of the Heisenberg spin chain Hamiltonian
$H_{\textrm{H}}$ above, we have: \[
H_{a}=\sum_{n=1}^{\left\lceil \frac{N-1}{2}\right\rceil }H_{\textrm{H}}^{(2n-1,2n)},\]
 and \[
H_{b}=\sum_{n=1}^{\left\lfloor \frac{N-1}{2}\right\rfloor }H_{\textrm{H}}^{(2n,2n+1)}.\]

Note that $H_{\textrm{H}}=H_{a}+H_{b}$. The Hamiltonians $H_{a}$
and $H_{b}$ consist of the couplings from the even spins to their
right neighbors and left neighbors respectively.

Our QCA evolution will consist of alternately evolving the lattice cells   under $H_{a}$
and $H_{b}$, using a technique called cell coloring. Each cell will
have two fields. The first field is a state register, consisting of
one qubit, which will hold the state of the spin represented by the
cell. The second field, called the active color register, will also
consist of a single qubit. Initially, the color register of each
cell $n$ is set to the value $n\bmod2$.

The QCA lattice used in this simulation is also one-dimensional, and
the neighbor set of each cell includes both the cell to the immediate
right, and the immediate left of the given cell. Let, $U_{n}'$ be
the Trotter step acting on the current cell state register and the
right neighbor state register. Using the first order approximation,
we have \[
U_{n}'=e^{-iH^{(n,n+1)}\Delta t/k}\]
 for an appropriate value $k$. It is also possible to use higher
order approximations.

The local update rule operator $U_{n}$ then consists of applying
the operator $U_{n}'$ if and only if the current cell's active color
register is set to one, and both left and right neighbors have their
color registers set to zero. The operator $V_{n}$ simply toggles the
active color register.

It should be clear that this QCA construction simulates the Heisenberg
spin chain. Moreover, by using an appropriate operator $U_{n}'$,
it is possible to simulate any Hamiltonian with nearest neighbor couplings
with this technique.

It is appropriate here to mention that one-dimensional spin structures
such as these can be efficiently simulated using classical computers
\cite{vidal:040502}. There are also efficient ways to calculate
the lowest energy eigenstates and eigenvalues using classical numerical
techniques such as the Density Matrix Renormalization Group (DMRG) method \cite{DMRGbook}. This, of course, also implies that the QCA presented
in this section can be efficiently solved and simulated classically.

However, this conclusion cannot be easily generalized to larger classes
of QCA. First, we note that we have used one-dimensional spin networks
here for expository purposes. From our constructions, it should be
clear that these QCA simulations generalize easily to higher dimensions.
On the other hand, no efficient classical simulation is known for
spin networks of dimension higher than one.

Also, it is not known whether arbitrary one-dimensional LUQCA can
be simulated efficiently classically. In fact, due to the universality
of one-dimensional LUQCA \cite{cpthesis}, this will not be the case
unless classical computers can efficiently simulate quantum systems
($BPP=BQP$), which is generally regarded as unlikely.

\subsection{Quantum Lattice Gases}

Quantum lattice gases have been studied for over a decade now \cite{boghosian98simulating,boghosian98,meyer96,yepez99quantum,Love:2005ve,lbm}.
In essence, they are the quantum analog of classical lattice gases.
The basic principles are the same in both the classical and quantum
cases: one starts with a discrete CA-based model that describes particles
on the lattice, and their movement. One can then take the \emph{continuous
limit} of such CA and show that in this limit, the behavior of the
CA mimics a well-known differential equation.

Taking the continuous limit of a classical CA is a well known procedure.
It involves giving the lattice a physical interpretation, where each
cell is thought to represent a point in space. The distance between
two adjacent cells is taken to be $\Delta x$ and each time step of
the CA is assumed to take $\Delta t$ time. One then takes the limit,
in a well prescribed manner, where $\Delta x\rightarrow0$ and $\Delta t\rightarrow0$.
There exist classical CA whose continuous limits represent gas diffusion,
as well as various other fluid dynamics \cite{cabook2}.

In the quantum case, Meyer \cite{meyer96}, and Boghosian and Taylor
\cite{boghosian98} give a construction of a \emph{quantum} lattice
gas whose continuous limit is the Schr\"{o}dinger equation for a
freely moving particle. We now show how any type of lattice gas can
be represented under the local unitary QCA model.

We begin by introducing the Quantum Walk QCA $Q_{W}$. This QCA models
multi-particle quantum walks on a lattice. Each cell is allowed to
have \emph{two} particles, in orthogonal states (these two states
can be thought of as orthogonal spins). The lattice can have any number
of particles in total.

The construction is as follows. The QCA $Q_{W}$ is one-dimensional.
Each cell has two single-qubit registers, called \emph{Up} and \emph{Down}.
Each register will represent the presence of a particle in the lattice
site, with the appropriate spin, by being in the state $\ket{1}$,
and the absence of the corresponding particle by being in the state
$\ket{0}$.

\begin{figure}
\begin{center}
\includegraphics{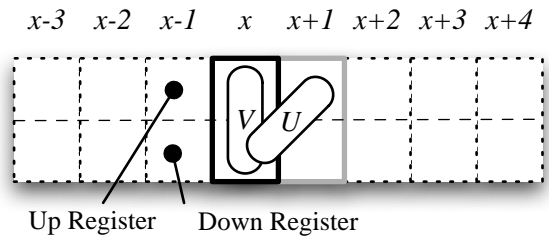} 
\end{center}
\caption{\emph{Quantum walk on a lattice.}\label{fig:QCAWalk}}
\end{figure}

The local update operator $U_{x}$ acts on the down register of the
current cell, and up register of the right neighbor, swapping the
two values. Operator $V$ operates on both fields of the cell with operator 
\[
V_{x}=\left(\begin{array}{cccc}
1 & 0 & 0 & 0\\
0 & q & p & 0\\
0 & p & q & 0\\
0 & 0 & 0 & \phi\end{array}\right),\]
 where $p,q\in\mathbb{C}$ satisfying $|p|^{2}+|q|^{2}=1$, $p\bar{q}+\bar{p}q=0$
and $|\phi|^{2}=1$.
The update rule is summarized in Figure \ref{fig:QCAWalk}.

The dynamics of this QCA are the same as the lattice gas described
by Boghosian and Taylor in \cite{boghosian98}.

Let $\Psi_{u}(x,t)$ and $\Psi_{d}(x,t)$ be the amplitude corresponding
to the presence of a particle with spin up and spin down respectively
in cell position $x$, at time $t$. Let $\Psi(x,t)$ be the total
amplitude corresponding to the presence of a particle in cell $x$
at time $t$, that is $\Psi(x,t)=\Psi_{u}(x,t)+\Psi_{d}(x,t)$. Then, we have that 
\[
\Psi_{u}(x,t+\Delta t)  =q\Psi_{u}(x-\Delta x,t)+p\Psi_{d}(x+\Delta x,t)\]
and
\[ \Psi_{d}(x,t+\Delta t)  =q\Psi_{d}(x+\Delta x,t)+p\Psi_{u}(x-\Delta x,t).
\]

We can proceed according to Boghosian and Taylor \cite{boghosian98},
and take the continuous limit of our QCA $\Delta x^{2}\rightarrow0$
and $\Delta t\rightarrow0$, using the Chapman-Enskog method \cite{cabook2}.
Doing so reveals that $\Psi(x,t)$ obeys the equation: \[
\frac{\partial}{\partial t}\Psi(x,t)=\frac{i}{2m}\frac{\partial^{2}}{\partial x^{2}}\Psi(x,t),\]
 which is the equation for a freely moving particle of mass $m=ip/q$
in one dimension.

Using the same construction techniques, we can also describe a freely
moving particle in two or three dimensions. We can construct QCA that
simulate other quantum lattice gases like the ones proposed in \cite{meyer96}.
Most, if not all, quantum lattice gases, whether single or multi-particle,
can be described as local unitary QCA.

This concludes our discussion on the expressive powers of the QCA model
presented here. In the next section we continue with a discussion
of how to take these mathematical models and implement them in quantum
hardware.

\section{Quantum Computation \label{sec:Quantum-Computation}}

In previous sections, we discussed how our unitary QCA can be used
to model physical systems, and how universal quantum computation
can be accomplished using only QCA primitives.
In this section we will look into bridging the gap by using QCA as
a theoretical framework for implementing quantum computation.

A clear advantage of working in the QCA model over quantum circuits,
in regards to physical implementations of quantum computation, is
that QCA make considerably fewer demands on the underlying hardware.
In particular, as opposed to direct implementations of quantum circuits,
the global evolution of the lattice in the QCA model does not assume
independent control over \emph{qudits}. Rather, all qudits are to
be addressed collectively in parallel. However, it should be noted
that the models of cellular automata described in this paper do not
explicitly address the issue of initialization. Any proposed physical
realization of the QCA must also describe the set of initial states
which are constructible. This may require some degree of non-global
control over a physical apparatus, such as having individual cells
initialized to a certain basis state, or it may require some interaction
with the environment, such as having the lattice cooled to a ground
state.

The QCA model also more closely resembles what is currently achievable
in several current quantum computer implementations. For example,
if qudits are represented by physical spins, and the control of such
spins is achieved through the use of magnetic pulses, as is the case
in NMR or ESR, then it is more reasonable to consider all spins as
being subjected to the same pulse sequences, rather than having the
ability to address spins individually.
The same can be said about many other physical quantum computer proposals.

In this section we will concentrate on implementing QCA on NMR, since
most of the groundwork for this implementation has already been laid
out.

\subsection{Colored QCA}

In Section \ref{sec:Modelling-Physical-Systems}, we considered cell
coloring as a useful QCA \emph{programming} technique. As with other
computation models, where a programming technique can be formalized
into its own subset model and then shown to be equivalent to the general
model (such as multi-track Turing machines), we can do the same with
colored QCA.

First, we will define the notion of a \emph{symmetric} transition
function for QCA. It is the quantum analog of symmetric CA, in which
the transition function depends only on the total number of neighboring
cells in each of the possible cell states. Essentially, a transition
update function is symmetric when it affects only the value of the
target cell in a manner which depends only on how many of the cell's
neighbors are in particular states, rather than on which state any
particular neighbor is in.

\begin{definition}
Given a QCA $Q=(L,\Sigma,\mathcal{N},U_{0},V_{0})$, we call the update
function $U_{0}:\left(\mathcal{H}_\Sigma\right)^{\otimes\mathcal{N}}
\rightarrow\left(\mathcal{H}_\Sigma\right)^{\otimes\mathcal{N}}$ symmetric if
it can be expressed as a collection of single-cell operations on cell $0$
controlled by the computational basis states of the neighborhood
$\mathcal{N}\setminus\{0\}$, and $U_{0}$ commutes with every operator
$\mathop{SWAP}_{x,y}$, which simply swaps the contents of cells $x$ and $y$, where
$x,y\in\mathcal{N}\setminus\{0\}$.  If $Q$ has a symmetric update function,
then we call $Q$ a symmetric QCA.
\end{definition}

Next, we wish to formalize the notion of a colored QCA.  For this model, we
will fix the neighborhood scheme to include only directly adjacent cells.
That is, $\mathcal{N}=\{x\in\mathbb{Z}^{d}:\left\|x\right\|_{1}\leq1\}$.
However, first we will define the set of permissible colorings of a lattice.

\begin{definition}
Given a lattice $L=\mathbb{Z}^{d}$ and a neighborhood scheme $\mathcal{N}$, we
define a \emph{correct} $k$-coloring for a lattice as a periodic mapping
$C:L\rightarrow\{0,1,\ldots,k-1\}$, such that no two neighboring cells in $L$
are assigned the same color.
\end{definition}

We may think of cell color as an inherent property of each cell.  However, it
may also be helpful to consider cell color as classical information which is
being stored with each cell in such a way that the local transition function
does not alter this information.  We can now finally give a definition for the
colored QCA.  Recall that the neighborhood scheme $\mathcal{N}$ is fixed.

\begin{definition}[CQCA]
A Colored QCA or CQCA is a 5-tuple $(L,C,\Sigma,\mathcal{U},c)$ consisting of
a lattice $L=\mathbb{Z}^{d}$, a correct $k$-coloring $C$, a finite set
$\Sigma$ of cell states, a sequence of $T$ symmetric unitary operators
$\mathcal{U}=\left(U_{0}^{(0)},U_{0}^{(1)},\ldots,U_{0}^{(T-1)}\right)$, with
$U_{0}^{(j)}:(\mathcal{H}_{\Sigma})^{\otimes\mathcal{N}}\rightarrow(\mathcal{H}_{\Sigma})^{\otimes\mathcal{N}}$,
and a sequence of $T$ colors $c=(c_0,c_1,\ldots,c_{(T-1)})$, labeled by
integers $0\leq c_j<k$.

The local transition operation consists of applying $U_{x}^{(j)}$ to each cell
$x$ with color $C(x)=c_j$ at time step $t=j+nT$, where $0\leq j<T$ and
$n\in\mathbb{Z}$.
\end{definition}
Note that since $C$ is a correct $k$-coloring, any two operators $U_{x}^{(j)}$
acting non-trivially on two cells of the same color at the same time will
commute.

CQCA can be simply considered as a shorthand for the cell coloring
technique we introduced in Section \ref{sec:Modelling-Physical-Systems}.
As such, it should be clear that CQCA are a subset of unitary QCA.

\begin{theorem} For every CQCA $Q$ there is a QCA $Q'$ that simulates
the same evolution exactly.
\begin{proof}
We may incorporate the color
information of each cell of the CQCA $Q$ within an additional color
register for each cell of the QCA $Q'$. Now, it suffices to add one
extra clock register to each cell, initialized to $0$. The update
operator $U_{x}$ simply applies $U_{x}^{(j)}$ conditional on both
$C(x)$ and the clock register of cell $x$ being set to $j$. In
order to ensure that $U_{x}$ commutes with its translations, we must
ensure that the colors of all the neighbors of $x$ are consistent
with the coloring $C$ before applying the appropriate operator.
Otherwise, $U_{x}$ should act as the identity operator. The read
operator $V_{x}$ simply increments the clock register, modulo $T$.
\end{proof}
\end{theorem}

What is more surprising is the converse result: that all unitary QCA
can be rephrased in the CQCA formalism.

\begin{theorem} \label{thm:QCA} For every QCA $Q$ there is a CQCA
$Q'$ that simulates the same evolution exactly.
\begin{proof}
Given the
QCA $Q=(L,\Sigma,\mathcal{N},U_{0},V_{0}),$ we will use the same
lattice $L$ and alphabet $\Sigma$. The neighborhood scheme for
the CQCA, $\mathcal{N}'$ is fixed  by definition. We also need to
provide a correct $k$-coloring of the lattice. To this end, it suffices
to provide a coloring $C$ with the property that that no neighborhood
$\mathcal{N}_{x}$ of $Q$ or $\mathcal{N}_{x}'$ of $Q'$ contains
two cells with the same color. Now, we need to construct a sequence
$\mathcal{U}$ of update operators. Note that single-qudit operations on $x$
and the controlled-$\mathop{NOT}$ operation targeting $x$ are symmetric
operations, since any two cells belonging to the same neighborhood have different colors, by construction. 
Now, given an implementation of the unitary update operation $U_{0}$
of $Q$ using single-qudit and nearest-neighbor controlled-$NOT$
operations, we can give a sequence of symmetric operations
which perform $U_{0}$ on a neighborhood $\mathcal{N}_{x}$ of a
cell $x$ of a specific color. By performing a similar sequence of
operations for each color in our coloring $C$, we effectively perform
$U_{x}$ for each cell $x$. Since each update operation $U_{x}$
commutes with the other update operations, we have effectively simulated
the update transition operation of $Q$. Finally, we can perform the
single-qudit operations $V_{x}$ on each cell.
\end{proof}
\end{theorem}

This last result is of major importance as it allows us to implement
any unitary QCA algorithm on a \emph{pulse-driven quantum computer},
as proposed by Lloyd \cite{lloyd93}, and further developed by Benjamin
\cite{benjamin1,Benjamin:2003oq} and others \cite{bw}. The scheme
involves using large molecules comprised of two or more different
species of spin-$\frac{1}{2}$ particles, arranged in repetitive structures,
such as crystals or polymers, to store the quantum data. It then evolves
the system using series of magnetic pulses that address all spins
of any one particular species.

To implement a given QCA in the pulse-driven computation model, we
first convert the QCA into one which uses a two-state alphabet. This
can be done by expanding each cell into $\left\lceil \log|\Sigma|\right\rceil $
cells to encode the states of $\Sigma$ with a binary alphabet, then
adjusting the neighborhood scheme $\mathcal{N}$ accordingly. We
then apply the construction in Theorem \ref{thm:QCA}. With this,
and the techniques of Lloyd \emph{et. al.}, it would be possible to
implement any QCA algorithm using NMR and an appropriate molecule.

We choose NMR and pulse driven quantum computing devices to show a
physical implementation of local unitary QCA. However, this should
not be taken to be the only possible implementation of QCA. There
are many other physical systems, like optical lattices \cite{Benjamin:2004wd},
cavity QED, among others \cite{kane1998,Khatun:2006bh}, that seem
better suited to implementing QCA, rather than the more traditional
quantum circuits.

\section{Algorithms \label{sec:Algorithms}}

We have seen two practical applications that can be achieved with
an implementation of QCA in the laboratory. First, there are numerous
physical systems that can be naturally simulated using the QCA model.
Second, one can also achieve universal quantum computation by simulating
quantum circuits on a QCA.

While these are both interesting and important applications of QCA,
a very important application in the future of QCA will be the development
and implementation of true, \emph{native}, QCA algorithms.

We saw in Section \ref{sec:Quantum-Circuits-and} how a quantum circuit
can simulate any QCA, and how a QCA can simulate any quantum circuit.
However, these simulations come at a cost of a linear-time slowdown
going in either direction. While this slowdown is not as important a concern in terms of asymptotic
complexity, in current laboratory conditions, \emph{any} source of
slowdown is to be avoided.

In the next section we analyze a problem that is particularly well-suited
to a natural solution using QCA, and we show how the tools that we
have developed thus far can be used effectively to provide an optimal
solution to the problem.

\subsection{Spin Signal Amplification Algorithm}

We present a description of the problem in simple abstract terms.
Suppose we want to \emph{amplify} the signal from a single spin-$\frac{1}{2}$
particle. That is, we have a single spin-$\frac{1}{2}$ particle,
and we want to create a large ensemble of spins whose bulk angular
momentum resembles the original spin in a particular basis. Note that
this is not cloning, since a basis needs to be set beforehand. Succinctly,
we want a unitary procedure $U$ that maps the state \[
\underbrace{\left(\alpha \ket{0} +\beta  \ket{1} \right)}_{\mbox{Amplified Spin}}\otimes\underbrace{\ket{0}^{\otimes N}}_{\mbox{Ancilla}}\]
 to the state \[
\alpha \ket{0}^{\otimes(N+1)}+\beta \ket{1}^{\otimes(N+1)},\]
 where $\ket{0}$ and $\ket{1}$ form the basis in which we wish
to \emph{amplify}. The main application of such an algorithm is to perform
a measurement in situations where bulk magnetization is needed in
order to achieve a detectable signal, such as with NMR. Hence, the
algorithm needs to be extremely efficient: the whole procedure needs
to be completed before decoherence can destroy the desired value.
The value $N$ will also need to be reasonably large, on the order
of $10^{7}$ or $10^{8}$, in order to get a reasonable signal in
NMR.

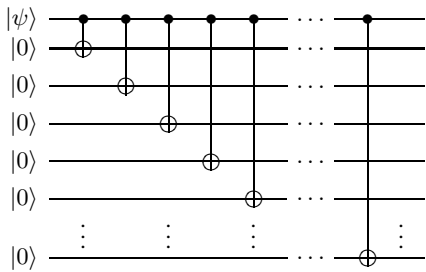
\begin{figure}
\begin{align*}
\Qcircuit @C=1em @R=.9em {
& \lstick{\ket{\psi}} & \ctrl{1} & \ctrl{2} & \ctrl{3} & \ctrl{4} & \ctrl{5} & \qw & \ldots & & \ctrl{7} & \qw & \qw \\
& \lstick{\ket{0}} & \targ & \qw & \qw & \qw & \qw & \qw & \ldots & & \qw & \qw & \qw \\
& \lstick{\ket{0}} & \qw & \targ & \qw & \qw & \qw & \qw & \ldots & & \qw & \qw & \qw \\
& \lstick{\ket{0}} & \qw & \qw & \targ & \qw & \qw & \qw & \ldots & & \qw & \qw & \qw \\
& \lstick{\ket{0}} & \qw & \qw & \qw & \targ & \qw & \qw & \ldots & & \qw & \qw & \qw \\
& \lstick{\ket{0}} & \qw & \qw & \qw & \qw & \targ & \qw & \ldots & & \qw & \qw & \qw \\
& & \vdots & & \vdots & & \vdots & & & & & \vdots \\
& \lstick{\ket{0}} & \qw & \qw & \qw & \qw & \qw & \qw & \ldots & & \targ & \qw & \qw }
\end{align*}\caption{\emph{A simple quantum circuit that implements $U$.} \label{fig:spin-circuit}}
\end{figure}

Figure \ref{fig:spin-circuit} shows a simple quantum circuit
solution. However, this circuit approach does have several shortcomings. First
and foremost, it requires individually addressing $N$ different spins.
For large $N$, in most laboratory conditions, this is not feasible.
Supposing that one could get around this first hurdle, one would still
need to perform $N$ independent gates before decoherence destroys
the data. Again, this is not likely to be feasible in most experimental
settings.

The QCA approach is simple, elegant, and optimally efficient.
In order to develop the algorithm we will make use of the colored
QCA developed earlier. We will use a two-color (black and white),
three-dimensional QCA. Since we are describing an algorithm that has
to be implemented in an actual physical setting, we will be using
a finite-sized workspace. In order to describe this workspace, we
will use three-state qudits for our cells: the logical states $\ket{{+}1}$
and $\ket{{-}1}$ will be used to denote the presence and the spin
of a spin-$\frac{1}{2}$ particle in the corresponding state, while
the state $\ket{0}$ will denote the absence of any particle. This
state $\ket{0}$ will be \emph{quiescent}, as defined in Section
\ref{sec:Local-Unitary-QCA}.

Every cell in the QCA is initialized to a state $\ket{0}$ except
for a perfectly cubic region of volume roughly $2N$. The cube will
have its cells initialized to the value $\ket{{-}1}$, except for
the top-front-left corner of the cube, whose value will be initialized to the state
$\ket{\psi}$ which we wish to amplify.
We will use this top, front, left portion of the cube as an
ancilla system.

As we are using a colored QCA, the neighborhood of each cell is
fixed to be the set of cells with Manhattan distance 1 from that cell.
Hence, each cell has only neighbors of the opposing color. We also
need to provide an update rule which is color-symmetric. For both
colors, the update rule is as follows. We apply a $NOT$ gate (which
maps $\ket{{+}1}$ to $\ket{{-}1}$, $\ket{{-}1}$ to $\ket{{+}1}$
and leaves the quiescent state $\ket{0}$ untouched) if and only
if the set of neighbors of a cell have values which sum to $-2$,
$-1$, or $0$. It can be shown that this update rule, when applied
repeatedly for $O\left(\sqrt[3]{N}\right)$ time steps, will achieve
the desired result. 

We can make a few simple observations about the algorithm. First,
as a native QCA algorithm, it does not require individual spin addressability.
The algorithm is optimally efficient, if we allow only the use of
local operations, in at most three dimensions.

It should also be noted that the problem of single-spin measurement
in NMR is generally considered to be a difficult one; the fact that
the exposition of the algorithm presented here is simple and succinct
is due to the development of the theoretical tools earlier in this
work.

It is important to add that it is possible to implement this algorithm
in solid state NMR by adapting some of the techniques presented above,
and applying some clever manipulations. For a full description of
this algorithm, including a discussion on physical implementation
see \cite{cp06,cp06-2,cpthesis}.

\section{Previous QCA Models \label{sec:Previous-QCA}}

In this section, we will present a number of other models of QCA that
have been developed, and we will relate them to our proposed model.

\subsection{Watrous-van Dam QCA}

The first attempt to define a quantized version of cellular automata
was made by Watrous \cite{watrous}, whose ideas were further explored
by van Dam \cite{vandamthesis}, and by D{\"u}rr, L{\^e}Thanh and
Santha \cite{durr-santha,durrsantha2}. The model considers a one-dimensional
lattice of cells and a finite set of basis states $\Sigma$ for each
individual cell, and features a transition function which maps a neighborhood
of cells to a single quantum state instantaneously and simultaneously.
Watrous also introduces a model of partitioned QCA in which each cell
contains a triplet of quantum states, and a permutation is applied
to each cell neighborhood before the transition function is applied.

\begin{definition}A Watrous-van Dam QCA, acting on a one-dimensional
lattice indexed by $\mathbb{Z}$, consists of a 3-tuple $(\Sigma,\mathcal{N},f)$
consisting of a finite set $\Sigma$ of cell states, a finite neighborhood
scheme $\mathcal{N}$, and a local transition function $f:\Sigma^{\mathcal{N}}\rightarrow\mathcal{H}_{\Sigma}$.
\end{definition}

This model can be viewed as a direct quantization of the classical
cellular automata model, where the set of possible configurations
of the CA is extended to include all linear superpositions of the
classical cell configurations, and the local transition function now
maps the cell configurations of a given neighborhood to a quantum
state. In the case that a neighborhood is in a linear superposition
of configurations, $f$ simply acts linearly. Also note that in this
model, at each time step, each cell is updated with its new value
simultaneously, as in the classical model.

Unfortunately, this definition allows for non-physical behavior.
It is possible to define transition functions which do not represent
unitary evolution of the cell tape, either by producing superpositions
of configurations which do not have norm 1, or by inducing a global transition
function which is not injective, and therefore not unitary.  In order to help
resolve this problem, Watrous restricts the set of permissible local
transition functions by introducing the notion of \emph{well-formed} QCA. A
local transition function is well-formed simply if it maps any configuration
to a properly normalized linear superposition of configurations. Because the
set of configurations is infinite, this condition is usually expressed in
terms of the $\ell_{2}$ norm of the complex amplitudes associated with each
configuration.

In order to describe QCA which perform unitary evolution, Watrous
also introduces the idea of a \emph{quiescent} state, which is a distinguished
element $\epsilon\in\Sigma$ which has the property that $f:\epsilon^{\mathcal{N}}\mapsto\epsilon^{\mathcal{N}}$.
We can then define a quiescent QCA as a QCA with a distinguished quiescent
state acting only on \emph{finite} configurations, which consist of finitely many non-quiescent states. It can be shown
that a quiescent QCA which is well-formed and injective represents
unitary evolution on the lattice. Also, note that this notion of a
quiescent state is slightly different than the one introduced in Section
\ref{sec:Local-Unitary-QCA}.

\begin{figure}
\begin{center}
\includegraphics{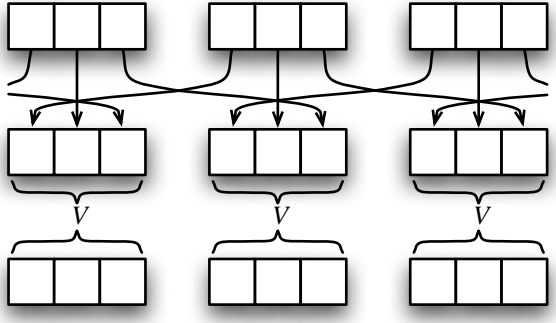}
\par\end{center}
\caption{\emph{Watrous Partitioned QCA.}}
\end{figure}

In order to construct examples of valid QCA in this model, Watrous also introduces a model of partitioned QCA, in which each cell consists of three quantum states, so that the set of finite states can be subdivided as $\Sigma=\Sigma_l\times\Sigma_c\times\Sigma_r$.
 Given a configuration in which each cell, indexed by $k\in\mathbb{Z}$, is in the state $(q_{k}^{(l)},q_{k}^{(c)},q_{k}^{(r)})$, the transition function of the QCA in one time step first consists of a permutation which brings the state of cell $k$ to $(q_{k-1}^{(l)},q_{k}^{(c)},q_{k+1}^{(r)})$ for each $k\in\mathbb{Z}$, then performs a local unitary operation $V_{k}$ on each cell.

Watrous shows that this model of partitioned QCA can be used to simulate
a universal quantum Turing machine with polynomial overhead.

\begin{figure}
\includegraphics{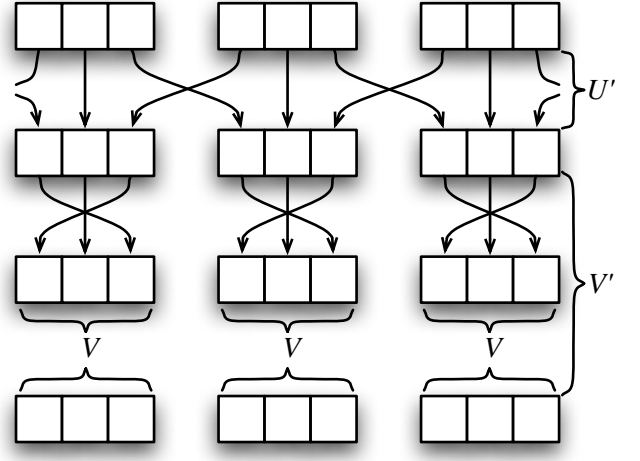}
\caption{\emph{Watrous QCA expressed as a Local Unitary QCA.} \label{fig:WQCA-as-unitary}}
\end{figure}

The partitioned QCA model given by Watrous can also be expressed in
the Local Unitary QCA model. First, suppose  $|\Sigma_l|=|\Sigma_r|$. If this is not the case, we can pad the smaller set with unused symbols so that both sets are of the same size.
Then, we separate the permutation
into an operation $P_{1}$ which operates on two consecutive cells,
mapping \begin{eqnarray*}
 & P_{1}: & (q_{k}^{(l)},q_{k}^{(c)},q_{k}^{(r)}),(q_{k+1}^{(l)},q_{k+1}^{(c)},q_{k+1}^{(r)})\\
 &  & \mapsto(q_{k}^{(l)},q_{k}^{(c)},q_{k+1}^{(l)}),(q_{k}^{(r)},q_{k+1}^{(c)},q_{k+1}^{(r)})\end{eqnarray*}
 followed by an operation $P_{2}$ which operates on a single cell,
mapping \[
P_{2}:(q_{k}^{(l)},q_{k}^{(c)},q_{k}^{(r)})\mapsto(q_{k}^{(r)},q_{k}^{(c)},q_{k}^{(l)}).\]
 Note that $P_{2}P_{1}$ performs the desired permutation, and also
that $P_{1}$ commutes with any lattice translation of $P_{1}$. Now,
we can express the Watrous partitioned QCA in our QCA model by setting
$U'=P_{1}$ and $V'=VP_{2}$, as shown in Figure \ref{fig:WQCA-as-unitary}.

\subsection{Schumacher-Werner QCA}

Schumacher and Werner \cite{sw04} take a different approach in the
definition of their model of QCA, working in the Heisenberg picture
rather than the Schr\"{o}dinger picture. They introduce a comprehensive
model of QCA in which they consider only the evolution of the algebra
of observables on the lattice, rather than states of the cell lattice
itself. By extending local observables of the cell lattice into a
closed observable algebra, the Schumacher-Werner model has a number
of useful algebraic properties. In this model, the transition function
is simply a homomorphism of the observable algebra which satisfies
a locality condition. Schumacher and Werner also introduce a model
of partitioned QCA called the \emph{Generalized Margolus Partitioned
QCA}, in which the observable algebra is partitioned into subalgebras.
This generalizes the Margolus scheme, as described in Section \ref{sec:Cellular-Automata},
in which the cell lattice itself is partitioned.

In order to avoid problematic issues dealing with observables over
infinite lattices, Schumacher and Werner make use of the \emph{quasi-local}
algebra. In order to construct this algebra, we first start with the
set of all observables on finite subsets $S\subseteq L$ of the lattice,
denoted $\mathcal{A}(S)$, and extend them appropriately into observables
of the entire lattice by taking a tensor product with the identity
operator over the rest of the lattice. The completion of this set
forms the quasi-local algebra.

In this setting, the global transition operator of a QCA is simply
defined as a homomorphism $T:\mathcal{A}(L)\rightarrow\mathcal{A}(L)$
over the quasi-local algebra which satisfies two specific properties.
First, a locality condition must be satisfied: $T(\mathcal{A}(S))\subseteq\mathcal{A}(S+\mathcal{N})$
for all finite $S\subseteq L$. Secondly, $T$ must commute with lattice
translation operators, so that the QCA is space-homogeneous. Now,
the QCA can be defined in terms of the lattice $L$, the neighborhood
scheme $\mathcal{N}$, the single-cell observable algebra, $\mathcal{A}_{0}$,
which takes the place of the alphabet, and the global transition operator
$T$.

The local transition operator of a QCA is simply a homomorphism $T_{0}:\mathcal{A}_{0}\rightarrow\mathcal{A}(\mathcal{N})$
from the observable algebra of a single distinguished cell $0\in L$
to the observable algebra of the neighborhood of that cell. Schumacher
and Werner show that a local homomorphism $T_{0}$ will correspond
uniquely to a global transition operator $T$ if and only if for each
$x\in L$, the algebras $T_{0}(\mathcal{A}_{0})$ and $\tau_{x}(T_{0}(\mathcal{A}_{0}))$
commute elementwise. Here, $\tau_{x}$ is a lattice translation by
$x$. The global transition operator $T$ given by $T_{0}$ is defined
by \[
T(\mathcal{A}(S))=\prod_{x\in S}T_{x}(\mathcal{A}_{x}).\]

Next, we will describe the Generalized Margolus Partitioned QCA. Schumacher
and Werner present this partitioned scheme as a method of producing valid reversible
QCA in their general model. In order to describe this scheme, we will proceed
according to the definition of a classical partitioned CA, as given
in Section \ref{sec:Cellular-Automata}.

We start with the $d$-dimensional lattice $L=\mathbb{Z}^{d}$, and
we fix the sublattice $S=2\mathbb{Z}^{d}$ as the set of cells of
$L$ with all even co-ordinates. We also fix the time period as $T=2$.
The block scheme, $\mathbf{B}$ is given as $\{B_{0},B_{1}\}$, which
is given as \[
B_{0}=\{(x_{1},x_{2},\ldots,x_{d})\in L:0\leq x_{j}\leq1,1\leq j\leq d\},\]
 which is simply a cube of size $2^{d}$ with corners at cells $\mathbf{0}=(0,0,\ldots,0)$
and $\mathbf{1}=(1,1,\ldots,1)$, and \[
B_{1}=B_{0}+\mathbf{1},\]
 which is simply a translation of the cube $B_{0}$.

Now, as in the regular Schumacher-Werner QCA model, we proceed in
the Heisenberg picture. For any block $B_{0}+s$, $s\in S$, we have
$2^{d}$ intersecting blocks from the partition $B_{1}+S$. For each
block $B_{1}+s'$ which intersects with $B_{0}+s$, there is a vector
$v\in\mathbb{Z}^{d}$ representing the translation taking $B_{0}+s$
to $B_{1}+s'$, so that $B_{1}+s'=B_{0}+s+v$. Indeed, these $2^{d}$
intersecting blocks may be indexed by the vectors $v$, which are
simply all vectors of $\mathbb{Z}^{d}$ whose entries are each $\pm1$.
Hence, we will set $B_{v}^{(s)}=B_{0}+s+v$.

For each block $B_{v}^{(\mathbf{0})}$, we will fix an observable
algebra $\mathcal{B}_{v}^{(\mathbf{0})}$ as a subalgebra of the observable
algebra $\mathcal{A}(B_{v}^{(\mathbf{0})})$ for the entire block.
Then, for each block $B_{v}^{(s)}$, the observable algebra $\mathcal{B}_{v}^{(s)}$
is simply the appropriate translation of $\mathcal{B}_{v}^{(\mathbf{0})}$.
Note that, in particular, the observable algebra for the block $B_{1}+s=B_{\mathbf{1}}^{(s)}$,
$\mathcal{A}(B_{\mathbf{1}}^{(s)})$, contains each of the observable
algebras $\mathcal{B}_{v}^{(s+\mathbf{1}-v)}$. In order for an assignment
of subalgebras to be considered valid, these subalgebras $\mathcal{B}_{v}^{(s+\mathbf{1}-v)}$
must commute and span $\mathcal{A}(B_{\mathbf{1}}^{(s)})$. This occurs
if and only if the product of the dimensions of these algebras is
$|\Sigma|^{2^{d}}$.

The transition function then consists first of an isomorphism \[
T_{0}^{(s)}:\mathcal{A}(B_{\mathbf{0}}^{(s)})\rightarrow\prod_{v}\mathcal{B}_{v}^{(s)},\]
 followed by the isomorphism \[
T_{1}^{(s)}:\prod_{v}\mathcal{B}_{v}^{(s+\mathbf{1}-v)}\rightarrow\mathcal{A}\left(B_{\mathbf{1}}^{(s)}\right).\]
 Note that since $T_{0}$ and $T_{1}$ are isomorphisms between observable
algebras of equal dimension, with an appropriate choice of basis,
they can be represented by unitary operators $U_{0}$ and $U_{1}$
which map vectors from a complex vector space to another complex vector
space of equal dimension. However, they do not represent local unitary
evolution, since these complex vector spaces are used to describe
two different quantum systems. For example, the Shift-Right QCA, which
was shown in Section \ref{sec:Local-Unitary-QCA} to be impossible
to implement using local unitary operators, can be constructed in the
Generalized Margolus Partitioning QCA model.

Fortunately, it is possible to simulate the Generalized Margolus Partitioning
QCA model within the Local Unitary QCA model by adding $2^{d}$ memory
registers to each cell corresponding to the subalgebras $\mathcal{B}_{v}$
in addition to a clock register indicating which of the two stages
of the transition function is being performed. The transition function
of the Local Unitary QCA simply swaps the contents of the data registers
of each cell with the appropriate memory registers before applying
the unitary operations corresponding to the desired isomorphisms.

\subsection{Other Models}

\citet{meyer96,meyer2}, \citet{boghosian98}, \citet{Love:2005ve}, among others explored the idea of using QCA as a
model for simulating quantum lattice gases. As classical CA are used
to model classical physical systems, it is natural to develop QCA
models which are capable of modelling quantum physical systems. In
order to simulate lattice gases, Meyer uses a model of QCA in which
each lattice cell is represented by a computational basis state in
a Hilbert space, and the set of states which a given cell can take
is replaced with a complex number representing the amplitude of the
basis state corresponding to that cell. In this regard, Meyer's QCA
modelling of lattice gases greatly differs from the one presented
here, and is not suitable as the basis for a more general model of
QCA.

Lloyd \cite{lloyd93} introduced a model of physical computation
based on a chain consisting of a repeating sequence of a fixed number
of distinguishable nuclear species. In this model, pulses are programmed which
are capable of distinguishing the species and performing nearest-neighbor
unitary operations. This model has been further developed by others
\cite{benjamin1,Benjamin:2004nx,Benjamin:2004eu}. It has been shown
that this model is sufficient for implementing universal quantum computation.

The model, sometimes referred to as \emph{`pulse-driven quantum computers'}, or Globally Controlled Quantum Arrays (GCQA), 
is different from QCA in that it allows for time-dependent evolution.
Still, they are closely related in their use of only space-homogeneous
update rules. For the sake of applying results pertaining to one model
to the other, it is also possible to argue that a pulse-driven quantum
computer is a degenerate case of a QCA where the update rule is applied
\emph{once}. Also, this physical scheme provides a natural platform
for implementing QCA.

\subsection{Comparison and Discussion}

In this section we have gone through a brief discussion of the major QCA models already in the literature today. While each has its own strengths, it is also true that each has weaknesses that are addressed by the LUQCA model.

We have seen how general Watrous-van Dam QCA have the problem of permitting ill-defined QCA. It is tempting to simply restrict attention to well-defined QCA in these models. However, deciding whether a given QCA is well-defined or not is a hard problem.

Even when a QCA is guaranteed to be well defined, as is the case for the partitioned Watrous-Van dam QCA, these allow for evolution that is not unitary and local, \emph{e.g.} shift-right automata as described  in Section \ref{sub:Shift-Right}.
This same criticism applies to the Schumacher-Werner model
of QCA.

Fortunately, it is possible to simulate any valid QCA in these models with local unitary QCA by adding ancillary space to each cell, in order to perform the necessary evolution in a unitary fashion.

Meyer's definition of  QCA, while being suitable for his purposes,  is not general enough to allow for all the behavior that is possible with local unitary QCA, \emph{i.e.} universal computation. Again, QCA in this model can easily be simulated by LUQCA.

Finally, we address globally-controlled quantum arrays. There are many similarities and connections between QCA and this model of computation. One important connection is how globally-controlled arrays can be used to implement QCA. However, these two models are quite distinct.
A GCQA is centered around the idea of doing computation on large arrays of simple quantum systems, without locally addressing them. GCQA divide their lattice of cells, or qudits, into \emph{subsets} each of which can be addressed collectively.

The first major distinction with QCA comes from the fact that sequences of pulses applied to these subsets of qudits are arbitrary, and do not necessarily follow a time-homogenous pattern.
The second, is that although Lloyd's construction is space homogenous, GCQA are not constrained in such a fashion. More recently, GCQA have been proposed that have less spatially homogenous structures \cite{fitzsimmons07}. 

As a model of computation one can say that QCA are more restricted than GCQA. At the same time, QCA are more than just a model of computation; they serve also as models of physical phenomena. It can be argued that QCA are, in a sense, a more \emph{fundamental} construct.

\section{Conclusions}

In this paper, we have presented a model of quantum cellular automata
based on local unitary operators. We have shown that it has distinct
advantages over previous cellular automata quantizations. In particular,
we have shown that given any LUQCA it is always possible to give an
efficient, low-depth quantum circuit that faithfully represents it.
This leads to the conclusion that any universal quantum computer could
implement an LUQCA efficiently.

More importantly, however, we have also shown that it is possible
to implement LUQCA in experimental setups that are arguably simpler
than traditional quantum circuit based algorithms: for instance, globally
addressing spins in NMR or ESR.
At the same time, we have shown that our model is universal for quantum
computation. We gave an explicit proof of an efficient simulation
of quantum circuits using a two-dimensional QCA and mentioned that
a proof of universality for one-dimensional LUQCA exists as well \cite{cpthesis}.
We also showcased the LUQCA as a modelling and simulation tool.

Finally, we have shown that QCA in previous models can be efficiently
translated into QCA within the model presented here. For example,
a universal QCA in previous models \cite{watrous,Shepherd:2006pd}
can easily become a universal QCA within the local unitary model (see
\cite{cpthesis} for an explicit construction). All of these facts suggest that the LUQCA is a very strong model.

The purpose of this paper has been to motivate, develop and showcase
a model of quantum cellular automata based on strictly local, translation-commuting,
unitary operators. It is our conjecture that the construction given
here is the most general of this form.

Ultimately, it is our hope that this paper serves to help unify the
several methods, results, and views surrounding QCA into one single,
cohesive paradigm.

\begin{acknowledgments}
The authors would like to thank Niel de Beaudrap and John Watrous for their invaluable comments on preliminary versions of this paper.
Research for this paper was supported in part by DTO-ARO, ORDCF, Ontario-MRI, CFI, MITACS, and CIFAR.
\end{acknowledgments}
\bibliography{qcabiblio}

\end{document}